\documentclass[twocolumn,showpacs,preprintnumbers,amsmath,amssymb,superscriptaddress]{revtex4}

\usepackage{graphicx}
\usepackage{dcolumn}
\usepackage{bm}
\usepackage{braket}
\usepackage{youngtab}
\usepackage{mathptmx}
\usepackage{url}

\usepackage{amsmath}
\usepackage{amsfonts}
\usepackage{mathrsfs}
\usepackage{amssymb}
\usepackage{revsymb}

\usepackage{color}

\begin{document}

\preprint{APS/123-QED}

\title{Detecting Hidden Differences via Permutation Symmetries}
\author{R. B. A. Adamson}
\affiliation{%
Centre for Quantum Information $\&$ Quantum Control and Institute
for Optical Sciences, Dept. of Physics, 60 St. George St.,
University of Toronto, Toronto, ON, Canada, M5S 1A7
}%
\author{P. S. Turner} 
\affiliation{Institute for Quantum Information Science,
 University of Calgary, 2500 University Drive NW, Calgary, AB, Canada, T2N 1N4}
\affiliation{Current address: Department of Physics, University of Tokyo, 7-3-1 Hongo, Bunkyo-ku, Tokyo, Japan, 113-0033}

\author{M. W. Mitchell}
\affiliation{ICFO - Institut de Ci\`{e}ncies Fot\`{o}niques, 08860
Castelldefels (Barcelona), Spain}

\author{A. M. Steinberg}
\affiliation{%
Centre for Quantum Information $\&$ Quantum Control and Institute
for Optical Sciences, Dept. of Physics, 60 St. George St.,
University of Toronto, Toronto, ON, Canada, M5S 1A7
}%

\date{\today}
\begin{abstract}
We present a method for describing and characterizing the state of
$N$ particles that may be distinguishable in principle but not in
practice due to experimental limitations.  The technique relies upon a careful treatment of the
exchange symmetry of the state among experimentally accessible and
experimentally inaccessible degrees of freedom.  The approach we
present allows a new formalisation of the notion of indistinguishability and can be
implemented easily using currently available experimental techniques.
Our work is of direct relevance to current experiments in quantum optics like
\cite{Mitchell2004,Eisenberg2005,Liu2006_1}, for which we provide a specific
implementation.
\end{abstract}

\maketitle

The predominant paradigm of quantum information science is the
qubit, a quantum two-level system.  This useful notion has allowed
many important concepts to be abstracted away from particular
physical implementations, revealing an underlying structure in the way
that information is manipulated and measured in quantum
mechanics.  Qubits are usually realised in some degree of freedom
of a physical system.  In many systems such as trapped ions and
nuclear spins, the physical particles are inherently separated and the
quantum statistical nature of the particles, be they bosons or
fermions, can safely be neglected.  In other systems including quantum
optics, degenerage atomic gases, optically trapped atoms, and
quasi-particles such as polaritons, the quantum
statistics of the particles can often play a role in the system's behaviour.  Sometimes the quantum statistical behaviour can be very
useful, as in the Hong-Ou-Mandel\cite{HOM1987} effect in quantum optical
systems which is often used to post-selectively implement interactions
between photons\cite{KLM2001}.

If a quantum information experiment is set up so that each particle is
uniquely different in some observable degree of freedom -- photons occupying different arms
of an interferometer, say, or ions in different locations in a trap
-- then the quantum statistical properties
of the particles generally do not play a role.  Characterization of
the quantum state of such systems proceeds according to the well-known
procedures of quantum state tomography\cite{MikeAndIke}.  The
influence of external, unobserved degrees of freedom can be accounted
for in this characterization and results in a density matrix displaying less-than-perfect coherence.

Recently, there have been several proposals and experiments
involving multiple photons occupying a single spatio-temporal mode and
two polarization modes\cite{Mitchell2004,Eisenberg2005,Bogdanov2004_1}. 
Such systems do not fit into the qubit paradigm and quantum statistics
plays an integral part in their behaviour.  The states of these systems are of enormous
interest\footnote{Some of these proposals refer to two multiply
  occupied spatial modes instead of polarization modes.  The two
  systems are formally equivalent and can be easily interconverted.} because they have been shown to exhibit phase
superresolution in interferometry\cite{Mitchell2004,Boto2000}, to be capable
of beating the diffraction limit in lithography \cite{Boto2000,Agarwal2007,Dangelo2001}, and
to open up new avenues in quantum imaging\cite{Bennink2004, Sanders1989}.  They have
also been proposed as a convenient qutrit useful in certain quantum
cryptography and quantum
information applications\cite{Bogdanov2004_1,Lang2004}.
While, to date, photon systems are the only ones where such states can
be created, recent developments
in optical lattices\cite{Folling2007} and elsewhere promise to
open up similar opportunities in other physical systems in the near future.

Because these states involve multiple occupancy of a single mode,
the quantum statistical nature of the particles is crucial to
understanding their behaviour.  Usually in considering
such states the formalism of creation and annihilation operators on
the field mode is used.  For example, the N00N state $\ket{N,0:0,N}$\cite{Boto2000} can be
written as
\begin{align}
\frac{1}{\sqrt{2(N+1)!}}\left({a_{1}^\dagger} ^N+ {a_{2}^\dagger}^N \right)\ket{0},
\end{align}
where the subscript indexes the distinct modes.
When such states are created experimentally, a central task is to
reconstruct a faithful characterization of the state from measurement statistics.

Ideally one would prefer to assume nothing about the source of the
quantum states, treating it as a `black box', and assume only that one
has a set of measurements that one is able to accurately perform on a
particular degree of freedom such as polarization.  The reconstruction
of the state from the measurements is called quantum state tomography
and it has been an essential tool in quantum state engineering, quantum
information science and quantum computing\cite{MikeAndIke}.  If the source produces an
indefinite number of photons then
continuous variable homodyne tomography methods can be extended to these
states\cite{Raymer2000}.  If the number of photons is
known, though, it is simpler to extend the quantum state tomography
techniques developed for qubits to
systems of multiply occupied modes, as was done for example by
Bogdanov \emph{et al}\cite{Bogdanov2004_2}.  

In their procedure one creates a basis of
states from creation operators for a single spatio-temporal mode and
the polarization modes that the state can occupy.  For the two-photon
case that they studied their basis states were $\left\{a_H^\dagger
a_H^\dagger, a_H^\dagger a_V^\dagger, a_V^\dagger
a_V^\dagger \right\}$, all taken to act on the vacuum.

There is a subtle assumption in writing the states this
way that runs counter to the purpose of quantum state tomography.  The
notation for the states \emph{assumes} that the raising operators act
on the same spatio-temporal mode, but this is not something that can
be experimentally verified from the polarization measurements
performed in the tomography.  Indeed since the experiment in
\cite{Bogdanov2004_2} involved combining different spontaneous
parametric downconversion sources into a single spatial mode there is
every reason to think that the different phase-matching conditions in
the two downconversion crystals would result in the raising operators
for each source acting on a somewhat different frequency-time mode.

If the two raising operators do indeed act on different
spatio-temporal modes then there can be a direct impact on
polarization measurements since the amplitudes for different
polarizations will carry the bosonic enhancement factors that one
obtains by multiplying raising operators on the same mode.  
In principle one could attempt to fully characterize the spatial and temporal degrees of
freedom to obtain the correct raising operator for each photon.  Such a full characterization
is technically very difficult, if not impossible.  Moreover, full
information about the spatio-temporal modes is likely not even desireable when it is 
ultimately polarization that is the degree of freedom of interest.  

What one would like is a `black box' tomography technique for
reconstructing the state in terms of polarization measurements only.  The
resulting description of the state ought to somehow include the influence of
all unobserved degrees of freedom on polarization measurements.  It
ought also to correctly predict the outcome of any
polarization measurement one wishes to perform so as to be considered `tomographically
complete'.  This paper develops and analyzes exactly such a technique.

To our knowledge the problem of charactizing the state of multiply
occupied modes has only arisen experimentally in photonic polarization
systems\footnote{During the preparation of this paper it came to our
  attention that a similar theoretical tomographic structure has been developed
  for spins in an unpublished manuscript by D'Ariano, Maccone and
  Paini\cite{DAriano2002}}. 
While we will concentrate on this specific realization, our method is
completely general and can be applied to any of the aforementioned
physical systems in which quantum statistics play a role, either
bosonic or fermionic.  In order to have the discussion that follows
reflect this generality we will define some technical language.  We
will call the information-carrying degree of freedom in such
systems the `visible' degree of freedom.  In the case of photonic
polarization systems such as in \cite{Bogdanov2004_2,Mitchell2004} this is polarization.  All
other degrees of freedom to which the apparatus is not sensitive we
call `hidden'.  The description of the state that
results from our state tomography procedure we label the `accessible' density matrix $\rho_\text{acc}$.  

While inspired by practical problems encountered in our attempts to
characterize quantum states, our approach
is interesting in its own right as an exploration of how adding
distinguishing information in experimentally inaccessible degrees of
freedom affects the quantum statistical properties of states.   

The key to our approach is to separate the state explicitly into
hidden and visible parts and to examine the constraints placed on
the visible degree of freedom by the quantum statistical requirements
on the whole state.  For photons, the bosonic statistics
require that the whole state be invariant under exchange
of all particle labels.  The exchange symmetry is more easily studied in
state notation rather than raising operator notation, so we will use
state notation throughout this paper.  This can be confusing at first because often in the literature
states are written in a way that does not make the exchange symmetry
explicit. For example, one might write the
polarization state of two photons as $\ket{HV}$, which is not
obviously exchange-symmetric as it must be for bosons.  In such a description the order of
the two labels implies the existence of a degree of freedom other
than polarization, say different spatial modes $a$ and $b$.  $a$ and
$b$ could be, for example, the distinguishable output angles of
downconverted photons in spontaneous parametric downconversion.  The
full bosonic state needs to be symmetric under exchange of both
spatial and polarization degrees of freedom, and would be written as 
$\ket{\psi_1}= \left(\ket{HV}\ket{ab}+\ket{VH}\ket{ba}\right)
/\sqrt{2}$.  Note that exchange of both the spatial and polarization
labels leaves the state invariant, but the state has the property
that one of the spatial modes, $a$ is always correlated with one
polarization $H$ and the other mode $b$ is always correlated with
polarization $V$.  In cases where the individual photon polarizations can be
treated as qubits because $a$ and $b$ are distinguishable paths, this
notation is redundant because no use is made of the permutation properties
of the whole state.  For this reason it is usually preferable to write the state as
$\ket{H_a V_b}$ which denotes the correlation between spatial and
polarization modes without making explicit the bosonic exchange
symmetry.  It should be understood that in all circumstances this way of writing the state
is simply a shorthand for $\ket{\psi_1}$.  

We emphasize this point because we wish to discuss situations where
the overall exchange symmetry of the state is important and the
notation of $\ket{\psi_1}$ becomes very useful.  There are
situations where the spatial modes in the above example are `hidden', that is to say they are
not resolved by the detection apparatus.  This might occur if the
photons were nearly collinear, but with a small angle between them.  A multimode collection system such as a lens focusing
onto a photodetector significantly larger than the optical
wavelength would have no means of distinguishing these two slightly
different spatial modes.  More generally, there could also be unresolvable
`hidden' time-frequency modes that can become occupied due to
uncorrected delays or dispersion.  Since the nanosecond-scale resolution of
most single-photon detectors is much longer than the femtosecond
timescale of pulsed experiments, different time-frequency modes are
generally not resolved by detectors.  In non-photonic systems there
are also myriad reasons why a given degree of freedom might be `hidden'
from experimental measurements.  When a hidden degree of freedom is
different for two particles we sometimes say that the
particles are distinguishable in principle but not in practice.

We can express our ignorance about the state of these `hidden' degrees
of freedom by tracing over them.  This leaves a density matrix observable only in
the visible degrees of freedom that we call the accessible density matrix:
\begin{align}
\rho_\text{acc}=\text{Tr}_\text{hid}\left[ \rho \right]\,.
\end{align}
For example, if in the state $\ket{\psi_1}$ the modes $a$ and $b$ cannot be resolved then we trace
over them to obtain the accessible density matrix
\begin{align}
\rho_\text{acc}&=\text{Tr}_\text{hid}\left[ \ket{\psi_1}\bra{\psi_1}\right]\\
&={\frac{1}{2}} \ket{HV}\bra{HV}+ {\frac{1}{2}} \ket{VH}\bra{VH}\,
\end{align}
This is a mixed state of polarization.  If the two photons
had occupied the same spatial mode so that the state was 
$\left(\ket{HV}\ket{aa}+\ket{VH}\ket{aa}\right)/\sqrt{2}$, then tracing
over the spatial degree of freedom would have yielded a pure
accesible density matrix in polarization
$\frac{1}{2}\left(\ket{HV}+\ket{VH}\right)\left(\bra{HV}+\bra{VH}\right)$.
Since these two situations yield different density matrices on the
polarization degree of freedom they can be distinguished by
polarization measurements alone.  The particular feature that
distinguishes them is the antisymmetric part, expressed as the population of the
singlet state $\left(\ket{HV}-\ket{VH}\right)/\sqrt{2}$.  The
singlet state projection makes up one element of the accessible
density matrix. It is a
measurable quantity \emph{even when the experimental apparatus cannot
tell the two photons apart}.  As we discuss extensively in \cite{Adamson2007}, for two photons the presence of an antisymmetric component of the
polarization state implies the existence of one or more unobserved
degrees of freedom that are different for the two particles and correlated in just the right way to result
in the correct bosonic exchange symmetry for the whole state.  This
shows that differences in the hidden degrees of
freedom may be inferred from measurements performed only on the
visible degrees of freedom.  

The remainder of this paper will examine how the accessible density
matrix can be calculated and measured for an arbitrary
number of particles and for a visible degree of freedom with an
arbitrary, finite number of levels.  In the first section we will begin by determining how
many elements are contained in a general accessible density matrix as
a function of the dimensionality of the visible degree of freedom and
the number of particles.
This determines both how many linearly independent measurements can be
made and how many numbers are needed to calculate all possible
expectation values on the visible degrees of
freedom.  The second section will put the discussion of the first
section on a firm group-theoretical footing.  In the third section we
examine how the theory applies to the case of three photon
polarizations.  In section 4 we discuss how the accessible
density matrix can be measured and work through a specific
numerical example with three photons.  Finally, in section 5 we discuss what
claims can be made about the indistinguishability of the particles from
a knowledge of the accessible density matrix.  

\section{The form of the accessible density matrix}\label{sec:den}

In this section we develop the structure of the accessible
density matrix and show how many independent measurements can be done
on a visible degree of freedom.  We will start by assuming a two-level
degree of freedom like photon polarization and then extend the result
to a $d$-level degree of freedom that could, for example, be the
Laguerre-Gauss spatial mode\cite{Lang2004} of photons.  

Our approach is to consider the Hilbert
space of the photons as a tensor product of a Hilbert space
describing the visible degrees of freedom and another describing
the hidden degrees of freedom.  Consider $N$ photon polarizations.  
Polarization transformations
$e^{i\left(x \sigma_x+y \sigma_y+z \sigma_z\right)}$ (where 
$\set{\sigma_x,\sigma_y,\sigma_z}$ are the Pauli matrices) give a realization of the 
group SU(2).   Since SU(2) acts irreducibly on the photon 
polarization, we can view the photons as spin one-half systems; that is, they transform according to the $j=\frac{1}{2}$ irrep of SU(2).  If all systems are
distinguishable in practice, so that each photon is in a separate mode
that can be experimentally distinguished (different rails of a
multi-rail interferometer, say), then the dimension of
the accesible space is the full dimension $2^N$ and the number of
accessible density matrix elements is $2^{2N}$.  This is the
familiar situation of quantum
state tomography as applied to photon polarization\cite{James2001}, trapped ions and
other qubit systems.  We would like to know the comparable number of density matrix elements
when the photons are not experimentally distinguishable because the
degrees of freedom that might distinguish the particles cannot be
resolved experimentally.  We note that in this case $2^{2N}$ provides an upper
bound on the number of elements in the accesible density matrix.

The following decomposition of the $N$-polarization Hilbert space will be
useful.  Unitary polarization operations acting on the whole state can
be decomposed into `angular momenta' $j$ (irreducible representations of
SU(2)) according to the well known Clebsch-Gordan series; for example
\begin{eqnarray}
\frac{1}{2}^{\otimes 2}
& = & 1 \oplus 0, \nonumber \\
\frac{1}{2}^{\otimes 3}
& = & \frac{3}{2} \oplus \frac{1}{2} \oplus \frac{1}{2}, \label{CGseries} \\
\frac{1}{2}^{\otimes 4}
& = & 2 \oplus 1 \oplus 1 \oplus 1 \oplus 0 \oplus 0.\nonumber
\end{eqnarray}
Notice that if $N>2$, certain $j$ values occur more than once; they
are said to have multiplicity.  However the largest $j$ always occurs only once,
since there is only one way to couple the spin-$\frac{1}{2}$ particles to maximum
$j=\frac{N}{2}$.  The states in the $\frac{N}{2}$ space are always totally
symmetric under permutation of the $N$ polarizations.  If they are indistinguishable 
\emph{in principle}, {\it i.e.} their hidden degrees of freedom are in the same state, 
then these totally symmetric visible states are the only ones available to the
whole state by the restriction that it have bosonic symmetry.  Since the
dimension of a spin $j$ space is $2j+1$, in this
case the dimension is $2\frac{N}{2}+1=N+1$ and the number of
accessible density matrix elements is $(N+1)^2$.  Previous tomography
schemes such as the one used in \cite{Bogdanov2004_1,Bogdanov2004_2}
worked under the tacit assumption
that the photons were indistinguishable in principle, and so described
the polarization only in terms of these $j=N/2$ states.

For experimentally distinguishable particles we see that the number of
density matrix elements grows exponentially as $2^{2N}$ with the
number of particles.  And as we have just shown, for particles indistinguishable in principle,
the number of elements grows polynomially as $(N+1)^2$ in the number of particles.
How does the number grow when the particles are distinguishable in
principle, but not in practice?  In this case, we
must trace out the hidden degree of freedom in order to express our ignorance
about them, but in doing so we are forced to erase the
ordering information of the $N$ systems.  This information is encoded
both in the phase between different terms in the Clebsch-Gordan
decomposition and, when multiplicity is greater than one, in how
population is distributed among the orthogonal eigenvectors of the multiplicity space; in
terms of operations, the unitary polarization transformations take
states with angular momentum $j$ to other states with angular momentum
$j$ in the same multiplicity space, while
permutations take states from one multiplicity space to a different
multiplicity space of the same $j$.

Sectors of states all carrying the same value of $j$ form $(2j+1)$ by $(2j+1)$ block-diagonal submatrices along
the main diagonal of $\rho_{\text{acc}}$.  SU(2) operations rotate
states within these blocks and permutations of the polarization labels
move population from one block to another with the same value of $j$.
\begin{equation}
\rho_{\text{acc}}=\left[ 
\begin{array}{cccccccc}
\ast & \ast & \ast & \ast & \multicolumn{4}{l}{ \mathrm{SU(2)}\,\mathrm{acts} } \\
\ast & \ast & \ast & \ast & \multicolumn{4}{l}{ \mathrm{within} } \\
\ast & \ast & \ast & \ast & \multicolumn{4}{l}{ \mathrm{blocks} } \\
\ast & \ast & \ast & \ast & & & & \\
\multicolumn{4}{c}{ \mathrm{S}_N\,\mathrm{acts} } & \ast & \ast & & \\
\multicolumn{4}{r}{ \mathrm{between}\nearrow } & \ast & \ast & & \\
\multicolumn{4}{r}{ \mathrm{blocks}\searrow } & & & \ast & \ast \\
& & & & & & \ast & \ast
\end{array}
\right]
\label{formofrhoacc}
\end{equation}

This explains why the highest $j$ space is symmetric --- the space has multiplicity one, and so must be invariant under permutations.

When we trace out the hidden degrees of freedom, coherences between
states of different $j$ as well as all information about the
state within the multiplicity spaces are destroyed, leading to a density matrix that is block
diagonal.  A consequence is that populations in multiple copies of the same $j$ are
averaged, yielding multiple copies of the same density submatrix.
Thus the accessible density matrix consists of only one independent
density submatrix for each $j$ in the Clebsch-Gordan decomposition
with zero coherence between submatrices.  The number of independent
accessible density matrix elements is therefore

\begin{align}
\sum_{j=0 \, \mathrm{or} \, \frac{1}{2}}^{\frac{N}{2}} (2j+1)^2 = {N+3 \choose 3}.
\label{qubitdimensions}
\end{align}
Thus the number of density matrix elements scales polynomially in the
number of particles, at least for two-level systems like
polarization.  

When the visible degree of freedom has $d$ distinct levels the
situation is completely analogous, with the Clebsch-Gordan series generalised to SU($d$).  
The space of $N$ $d$-level systems decomposes into irreps $\lambda$
of SU($d$) where now the label
$\lambda=(\lambda_1,\lambda_2,\cdots,\lambda_d)$; $\sum_i \lambda_i=N$, $\lambda_i \geq \lambda_{i+1}$ 
is a regular partition of $N$ (a Young diagram).  Of course, if the systems are experimentally distinguishable
then the entire $d^N$ dimensional Hilbert space is
accessible and the number of accessible density matrix elements is
$d^{2N}$ which gives the (exponential) upper bound.  The irrep
$(N,0,0,\cdots,0)$, (analogous to highest $j$ in the SU(2) case), occurs only once in the decomposition of the
Hilbert space and so is always symmetric under permutations.  The dimension of
this (and indeed any SU($d$)) irrep is given by the Weyl character formula~\cite{Weyl}
\begin{align}
\mathrm{dim}(\lambda) = \prod_{1\leq i < j\leq d} \frac{\lambda_i - \lambda_j + j - i}{j - i}. 
\end{align}
If the qudits are indistinguishable in principle, then again they are
restricted to the totally symmetric subspace with $\lambda_1=N$ and
all other $\lambda_j=0$; the Weyl formula gives
\begin{align}
\prod_{j=2}^d \frac{N+j-1}{j-1} = {N+d-1 \choose N} 
\end{align}
for the dimension, so the number of accessible density matrix elements
is ${N+d-1 \choose N}^2$.

The unitary and permutation group actions are the same as in the SU(2)
case.  SU($d$) acts within irrep spaces $\lambda$ and S$_N$ acts
`across' multiplicities.  When the distinguishing degrees of freedom
are hidden, the ordering information of the systems is lost and the
permutation group action is trivialised, leaving only one `copy' of
each SU($d$) irrep space for each $\lambda$.  The dimension of the accessible
space is therefore\footnote{This can be proved using the Cauchy formula for the general linear group.  T. A. Welsh, private communication.}
\begin{align}
\sum_\lambda \prod_{1\leq i < j\leq d} \frac{\lambda_i - \lambda_j + j - i}{j - i}
= {N+d^2-1 \choose N},
\end{align}
and is always a polynomial in $N$. 

\section{Group Theoretical Construction}\label{app:}
Here we will construct explicitly the most general totally symmetric
state of a system of particles with both visible and hidden degrees of
freedom, which we use to justify the claims made in the last section.  Let the Hilbert spaces for these two be denoted
$\mathscr{H}^\mathrm{vis}$ and $\mathscr{H}^\mathrm{hid}$,
respectively.  Assuming that there are $N$ particles, the same
permutation group S$_N$ acts on both of these spaces.  Decompose each
space into irreps $\lambda$ of S$_N$ and consider their tensor
product, the space of all available states:
\begin{align}
\mathscr{H} = \left( \bigoplus_{\lambda,m} \mathscr{H}^\mathrm{vis}_{\lambda,m} \right) \otimes \left( \bigoplus_{\lambda',m'} \mathscr{H}^\mathrm{hid}_{\lambda',m'} \right),
\end{align}
where $m$ labels the multiplicity of irrep $\lambda$.  Let $\mu$ index
an orthonormal basis for each irrep space $\mathscr{H}_{\lambda,m}$;
the basis states are labelled
\begin{align}
\ket{\lambda m \mu}_\mathrm{vis} \ket{\lambda' m' \mu'}_\mathrm{hid},
\end{align}
where $m=1,2,\cdots,\text{mult}\mathscr{H}_\lambda$ runs over the
multiplicity of irrep $\lambda$ in the Hilbert space, and
$\mu=1,2,\cdots,\text{dim}\mathscr{H}_\lambda$ runs over the dimension
of irrep $\lambda$. For readers familiar with Schur-Weyl duality, $m$
indexes a basis for an irrep $\lambda$ of the unitary group action on
each particle, and $\mu$ indexes a basis for an irrep $\lambda$ of the
permutation group action S$_N$.  The fact that the same irrep label
can be used for both group actions is why they are `dual'.

Now the problem of finding totally symmetric states in $\mathscr{H}$
is a coupling problem, completely analogous to coupling angular
momentum states to arrive at states of angular momentum zero.  In
fact, it can be shown from the rules for tensor products of Young
diagrams that the totally symmetric irrep $\lambda=(N)$ of S$_N$ only
occurs in a tensor product $\lambda\otimes\lambda'$ if
$\lambda'=\lambda$, and moreover that $(N)$ only occurs once,
\textit{i.e.} it has multiplicity one~\cite{Hammermesh}.  The analogy
is that the spin zero irrep of the rotation group only
occurs in the tensor product $j\otimes j'$ if $j'=j$, and it occurs
only once, \textit{i.e} in order to couple two angular momenta to
$j=0$, we know the two angular momenta must be equal.  Note also that
$(N)$ is always one dimensional, so there is one totally symmetric
state for each $\lambda$, unique up to multiplicity.

Given an irrep $\lambda$ and two multiplicity sectors $m,m'$ in
$\mathscr{H}$, this unique totally symmetric (unnormalised) state
$\ket{\lambda m m'}$ is an equally weighted superpositon of the states of
each factor in the tensor product
\begin{align}\label{eq:ts}
\ket{\lambda m m'} \equiv \sum_{\mu=1}^{\text{dim}\mathscr{H}_\lambda} \ket{\lambda m \mu}_\mathrm{vis} \ket{\lambda m' \mu}_\mathrm{hid},
\end{align}
(which is a state on the combined space, not to be confused with the uncombined visible and hidden states, despite the fact that they both have three labels).  The most general totally symmetric pure state in $\mathscr{H}$ is therefore an arbitrary linear combination of these:
\begin{align}
\ket{\psi_N} = \sum_\lambda \sum_{mm'} C^\lambda_{mm'} \ket{\lambda m m'}.
\end{align}

The same analysis goes through for totally antisymmetric states.  The
unique coupling is $\lambda\otimes\bar\lambda$, where $\bar\lambda$ is
the irrep conjugate to $\lambda$.  There is a restriction, however,
given by the dimension of the Hilbert space for each particle, which
is again encoded in the rules for Young diagrams.  For example, there
is no totally antisymmetric state of three indistinguishable spins.

Now we can define what we mean by distinguishable and
indistinguishable.  Expand $\ket{\psi_N}$ in the physical basis of $N$
particles.  Those states in the expansion where the hidden degrees of
freedom for all $N$ particles are in the same state are \emph{indistinguishable} in
principle.  This hidden state is totally symmetric by definition, and by the coupling mentioned above it follows that the visible
state must also be symmetric.  Since $(N)$ is one dimensional, there is only one term
in the sum over the basis indexed by $\mu$ above, and the total state
is separable across the hidden and visible subspaces.  Thus, tracing
out the hidden space does not alter the visible state, and since it
can only lie in $(N)$, the accessible density matrix is restricted to
the totally symmetric subspace, as expected.

Those states in the expansion where the hidden degrees of freedom for
all $N$ particles are in distinct orthogonal states are \emph{distinguishable} in principle.  There is a
large amount of entanglement across the hidden and visible subspaces.
If the hidden modes are inaccessible in practice, then we arrive at
the accessible density matrix by tracing out the hidden modes.  Using
Eq.(\ref{eq:ts}), one finds
\begin{align}
\rho_\mathrm{acc} 
&= \mathrm{Tr}_\mathrm{hid} \left[ \ket{\psi_N}\bra{\psi_N} \right] \\
&= \sum_{\sigma \ell \nu} \bra{\sigma\ell\nu}\sum_{\lambda\kappa mm'nn'} C^\lambda_{mm'} C^{\kappa\ast}_{nn'} \ket{\lambda m m'} \bra{\kappa n n'} \ket{\sigma\ell\nu}_\mathrm{hid} \\
&= \sum_{\lambda m m'} \rho^{\lambda}_{mm'} \sum_\mu \ket{\lambda m \mu}_\mathrm{vis} \bra{\lambda m' \mu}.
\end{align}
One therefore concludes that $\rho^{\lambda}_{mm'} = \sum_n
C^\lambda_{mn} C^{\lambda\ast}_{m'n}$ affords the only freedom in the
accessible density matrix, giving only one value per irrep $\lambda$
and pair of multiplicity indices $m,m'$.  The trace erases coherences
between different $\lambda$ sectors on account of those sectors being
orthogonal.  We also see that the equally weighted average over $\mu$
which was necessary for total symmetry has destroyed any independence
between the multiplicity spaces --- we get the same copy of the
$\lambda$ submatrix for all $\mu$, and so we effectively have one
submatrix for each $\lambda$.  From the point of view of accessible
measurements the state space has `collapsed', although if the
particles were distinguishable in one of the hidden degrees of
freedom, then the ability to measure that degree of freedom would
restore the Hilbert space to its full size.  

The measurement of the accessible density matrix elements
$C^\lambda_{mn}$ allows one to infer the existence of hidden
differences among the particles making up the state.  To see
this, consider that the hidden and visible spaces must both transform under
the same permutation group S$_N$.  If we decompose the visible and
hidden spaces separately under this common group action, we arrive at
visible states labelled by S$_N$ irreps $\lambda$ and hidden states
labelled by S$_N$ irreps $\lambda'$. Again, coupling visible and hidden
states to make totally symmetric states is completely analogous to
coupling angular momentum states to make angular momentum
$j=0$.  It follows that $\lambda'$ must equal $\lambda$.  Thus, if a visible state is measured
to be in a state of permutation symmetry $\lambda$ that is not totally
symmetric, one can infer that there existed a hidden state of
permutation symmetry $\lambda$ to which it was coupled, implying
in turn the presence of multiple orthogonal states for the
hidden degrees of freedom.  These hidden differences serve to
make the the photons distinguishable and explain why the
coherences between different $\lambda$ ($j$ for SU(2))
disappear when the hidden states are traced, simply because states of
different $\lambda$ are orthogonal.

\section{Example: The accessible density matrix for three photon
  polarizations}

To make the discussion of the previous sections more concrete we will
focus on the particular example of three photon polarizations.  This
example is experimentally relevant to previously published work from our group on
N00N states\cite{Mitchell2004}, and to ongoing work on making other states in the same
three-photon polarization Hilbert space.

The Clebsch-Gordan decomposition for three spin-$\frac{1}{2}$
particles was given in Eq. (\ref{CGseries}).
We can explicitly write out the states of this decomposition. Each
state is labeled by a pair of angular momentum quantum numbers $j$ and
$m$.  The $j=3/2$ states that are completely symmetric under permutations are:
\begin{align}
&\ket{3/2,\,3/2}=\ket{HHH}\,,\\
&\sqrt{3}\ket{3/2,\,1/2}=\ket{HHV}+\ket{HVH}+\ket{VHH}\,,\\
&\sqrt{3}\ket{3/2,\,-1/2}=\ket{VVH}+\ket{VHV}+\ket{HVV}\,,\\
&\ket{3/2,\,-3/2}=\ket{VVV}\,.
\end{align}
While the $j=1/2$ space has multiplicity two.  The two spaces are
spanned by
\begin{align}
&\sqrt{6}\ket{1/2,\,1/2}_1=\ket{HHV}+\ket{HVH}-2\ket{VHH},\label{s2m1/2}\\
&\sqrt{6}\ket{1/2,\,-1/2}_1=\ket{VVH}+\ket{VHV}-2\ket{HVV},\label{s2m-1/2}
\end{align}
and
\begin{align}
&\sqrt{2}\ket{1/2,\,1/2}_2=\ket{HHV}-\ket{HVH},\label{s1m1/2}\\
&\sqrt{2}\ket{1/2,\,-1/2}_2=\ket{VVH}-\ket{VHV}.\label{s1m-1/2}
\end{align}
$\ket{1/2,\,1/2}_1$ transforms into $\ket{1/2,\,1/2}_2$ under
permutation operations and $\ket{1/2,\,-1/2}_1$ transforms into
$\ket{1/2,\,-1/2}_2$ in exactly the same way.
However, polarization measurements cannot distinguish $\ket{1/2,\,1/2}_1$ from
$\ket{1/2,\,1/2}_2$ or $\ket{1/2,\,-1/2}_1$ from
$\ket{1/2,\,-1/2}_2$.  All they can do is determine
the average of the two-by-two density matrix over the space spanned
by  $\ket{1/2,\,1/2}_2$
and $\ket{1/2,\,1/2}_1$ and the density matrix over the space spanned
by $\ket{1/2,\,-1/2}_2$ and $\ket{1/2,\,-1/2}_1$.  From the point of view
of polarization measurements, the information contained in the two spaces collapses
into a single effective $j=1/2$ sector of $\rho_{\text{acc}}$. 

The accessible density matrix contains ${3+3 \choose 3}=4^2+2^2=20$ elements.  When
distinguishing information is hidden, the best characterization
of the state of three photon polarizations is the determination of these 20 elements.

\section{Measuring the accessible density matrix}

\begin{figure}
  \centerline{
   \mbox{\includegraphics[width=\columnwidth]{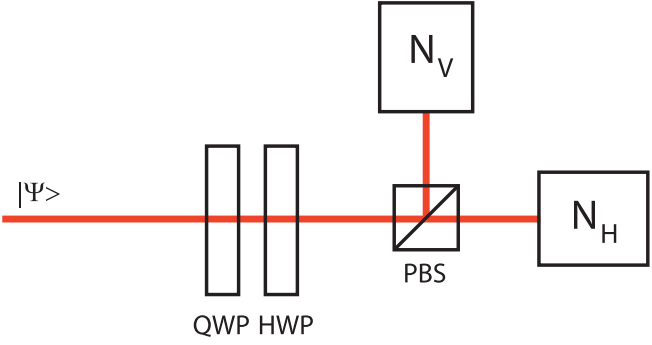}}
  }
  \caption{Apparatus for measuring the accessible density matrix for $N$
    photon polarizations.  The
  state $\ket{\Psi}$ is sent into a quarter waveplate (QWP) and half
  waveplate (HWP) followed by a polarizing beamsplitter (PBS).  Number
  resolving photon counters count the number of vertical photons $N_V$
  and the number of horizontal photons $N_H$.}
  \label{apparatus}
  \end{figure}

We have shown that elements of the accessible density matrix offer the most complete
description of the state of $N$ particles when one degree of freedom of
the particles is visible and others are hidden.  It is not clear from
our discussion so far that it is possible to measure $\rho_\text{acc}$
using available experimental tools.  In this section we will show that
in the case of polarization it is indeed possible to measure
$\rho_\text{acc}$ with a simple experimental device.  This device,
shown in figure \ref{apparatus}, involves four
different optical elements, a quarter wave-plate, a
half-waveplate, a polarizing beamsplitter (PBS) and number-resolving
photon counters (such as the one demonstrated in \cite{Achi2005}) at
the two ports.   The multiphoton polarization state passes through the
two waveplates and on to the PBS where it is split into $H$ and $V$
components.  The number of photons in each port is then measured with
the number-resolving single photon detector.  Such a device is
completely analogous to the Stern-Gerlach device for
measuring spin projections.

If there are $N$ photons in the state then there are $N+1$ different
ways that these can split between the $H$ and $V$ ports of the
polarizing beamsplitter.  When the photons leaving each port are
counted, the measurement implemented will be a convex sum of projectors
onto all states having that number of horizontal and vertical photons.
For example, 
${\mathbf P}_N=\ket{HH\cdots H}\bra{HH\cdots H}$,
${\mathbf P}_{N-1}=\left(\ket{VH\cdots H}\bra{VH\cdots H}+\ket{HV\cdots H}\bra{HV\cdots H}+\ldots\right)$, 
and so on.  In angular momentum language ${\mathbf P}_N$ is a pure projector onto the state $j=N/2$,
$m=N/2$. By changing the angles of the waveplates one can `orbit' this
measurement in the $j=N/2$ space thereby obtaining all the density
matrix elements lying in the $j=N/2$ space.  All $j=N/2$ matrix
elements can be determined by measuring only rotated versions of
${\mathbf P}_N$.  ${\mathbf P}_{N-1}$ is a convex sum of projectors onto all
states with $m=N/2-1$.  Since the $j=N/2$, $m=N/2-1$ projection can be
determined from rotated versions of ${\mathbf P}_N$, we can subtract
this part from ${\mathbf P}_{N-1}$ leaving a projector with support
only in the $j=N/2-1$ subspace.  By changing the waveplate angles
one can use this reduced operator to completely characterize the $j=N/2-1$ space. One
can then subtract the $j=N/2$ and $j=N/2-1$ terms from the $m=N/2-2$
operator, and so on.  In this way all the terms in the
accessible density matrix can be measured.

It should be noted that this is not the only way to measure the
accessible density matrix.  In fact any set of linearly independent
measurements equal in number to the number of accessible density
matrix elements as calculated from equation \ref{qubitdimensions} can be used to reconstruct the accessible density
matrix.  Standard inversion techniques such as linear inversion,
maximum likelihood fitting\cite{James2001} and convex maximum
likelihood fitting\cite{Kosut2004} can be used to obtain $\rho_\text{acc}$
from an experimental dataset.

\subsection{Three-photon example}
In this section we work through an example of how our techniques might
be applied in a three-photon polarization experiment.  We
consider an experimental situation similar to the one used to
create the state
$\frac{1}{\sqrt{2}}\left(\ket{3_H,0_V}+\ket{0_H,3_V}\right)$ in
reference \cite{Mitchell2004}.
There three photons were combined on a beamsplitter and post-selection was
used to isolate those instances where all three photons left from
the same port of the beamsplitter.  If all the photons were
indistinguishable, and each photon was set to the correct
polarization then by this procedure the entangled state 
$\ket{3,0:0,3}=\frac{1}{\sqrt{2}}\left(\ket{3_H,0_V}+\ket{0_H,3_V}\right)$
would have been.  This comes about because of the state is factorizable in raising
operators through the relation:
\begin{align}
\left(\right. {a^\dagger_H}^3+ & {a^\dagger_V}^3 \left. \right)\notag\\
&=\left(a^\dagger_H+a^\dagger_V\right)\left(a^\dagger_H+e^{ 2
  \pi i/3} a^\dagger_V\right)\left(a^\dagger_H+e^{4 \pi i/3}
a^\dagger_V \right) \label{noondecomposition}
\end{align}
Note that the right side is a product of polarization raising
operators all acting on the same spatio-temporal mode.

In that experiment, however, two of the photons were produced by a
spontaneous parametric downconversion process and the third was
produced by an attenuated laser pulse.  It is to be expected
that these different sources might produce photons with hidden
differences in their time-frequency wavefunctions.  In
principle such differences can be reduced by filtering, but
let us suppose that filtering is insufficient, resulting in the
mode of the third photon having only a 50\% overlap with
the mode of the other two photons, which are identical to one another.  We can model
this by replacing the raising operators in the third bracket in
equation \ref{noondecomposition} with operators
$c^\dagger_{H/V}=\frac{1}{\sqrt{2}}\left(a^\dagger_{H/V}+b^\dagger_{H/V}\right)$
  where $b^\dagger$ is a creation operator for a mode $b$ orthogonal
  to the mode of $a$ for which $a^\dagger$ is the raising
  operator.  The $50\%$ overlap is chosen here to keep the
  calculation simple.  For a more general situation one can
  repeat the analysis using $c^\dagger_{H/V}=\cos{\theta}
  a^\dagger_{H/V}+\sin{\theta} b^\dagger_{H/V}$ where
  $\theta$ is an angle parametrizing the degree of overlap.

Inserting this substitution into equation
\ref{noondecomposition}, we obtain the expression
\begin{align}
\frac{1}{\sqrt{11}} & \left( a^\dagger_H  +a^\dagger_V
\right)\left(a^\dagger_H+e^{ 2 \pi  i/3} a^\dagger_V\right) \left(c^\dagger_H+e^{4 \pi i/3} c^\dagger_V\right) \notag \\
=&\frac{1}{\sqrt{11}}\left[
{a^\dagger_H}^3+{a^\dagger_V}^3\right]+\frac{1}{\sqrt{22}}\left[ \right. {a^\dagger_H}^2
  b^\dagger_H
  +  a^\dagger_H a^\dagger_V b^\dagger_H  \left(1+e^{2\pi i /3}\right)\notag \\
&+{a^\dagger_V}^2 b^\dagger_H e^{2\pi i/3} + {a^\dagger_H}^2 b^\dagger_V e^{4
    \pi i/3}+a^\dagger_H a^\dagger_V b^\dagger_V \left(1+e^{4 i\pi /3}\right)\notag \\ 
&+{a^\dagger_V}^2 b^\dagger_V \left. \right] \label{fullstate}
\end{align}
Our goal is to write the state as an accessible density matrix, purely in
terms of polarization measurements that can be done without
knowing anything about the differences between the hidden
time-frequency degrees of freedom.  To arrive at such an
expression we will need to trace over the orthogonal hidden
modes $a$ and $b$.  To do so the expression must first be
rewritten in a first-quantized notation that clarifies the
imposed separation between the hidden and visible degrees of freedom.  The full
expression is too long to write here, but rewriting one of the
terms should be enough to give a feel for the calculation.

Consider the term $a^\dagger_H  a^\dagger_V b^\dagger_V$.  In
rewriting this in first-quantized form we need to use tensor
products of state vectors on the hidden and visible degrees of
freedom, keeping in mind that the indistinguishability of the
particles means that any one of the three can be in mode $b$.
We write it as follows:

\begin{align}
a^\dagger_H & a^\dagger_V b^\dagger_V= \notag \\ 
&\frac{1}{\sqrt{6}}\left[\right. \left(\ket{HVV}+\ket{VHV}\right)\ket{aab}+\left(\ket{HVV}+\ket{VVH}\right)\ket{aba}
  \notag\\
&+\left(\ket{VHV}+\ket{VVH}\right)\ket{baa} \left.\right]
\end{align}

The trace over the hidden degrees of freedom produces an incoherent sum over the density matrices
for each bracketed polarization state since $\ket{baa}$,
$\ket{aab}$ and $\ket{aba}$ are orthogonal.  The resulting accessible
density matrix describing this term is 

\begin{equation}
\rho_{\text{acc}}=\left[ 
\begin{array}{cccccccc}
0 & 0 & 0 & 0 & \multicolumn{4}{l}{ } \\
0 & 0 & 0 & 0 & \multicolumn{4}{l}{ } \\
0 & 0 & 2/3 & 0 & \multicolumn{4}{l}{ } \\
0 & 0 & 0 & 0 & & & & \\
\multicolumn{4}{c}{  } & 0 & 0 & & \\
\multicolumn{4}{r}{ } & 0 & 1/6 & & \\
\multicolumn{4}{r}{ } & & & 0 & 0 \\
& & & & & & 0 & 1/6
\end{array}
\right]
\end{equation}

In the same way we can obtain the accessible density matrix for the entire state in Equation
\ref{fullstate}.

\begin{widetext}
\begin{equation}
\rho_{\text{acc}}=\left[ 
\begin{array}{cccccccc}
0.3636 & 0 & 0 & 0.3636 & \multicolumn{4}{l}{ } \\
0 & 0 & 0 & 0 & \multicolumn{4}{l}{ } \\
0.0 & 0 & 0 & 0 & \multicolumn{4}{l}{ } \\
0.3636 & 0 & 0 & 0.3636 & & & & \\
\multicolumn{4}{c}{  } & 0.0682 & -0.0341-0.0590i & & \\
\multicolumn{4}{r}{ } & -0.0341+0.0590i & 0.0682 & & \\
\multicolumn{4}{r}{ } & & & 0.0682 & -0.0341-0.0590i \\
& & & & & & -0.0341+0.0590i & 0.0682
\end{array}
\right]
\label{theorydm}
\end{equation}
\end{widetext}
Note that the partial distinguishability of the third photon results in
$27\%$ of the population being in the $j=1/2$ spaces instead of
the $j=3/2$ spaces.  The fidelity\cite{Jozsa1994} of this state
to the desired state $\ket{3,0:0,3}$ is 0.7273.  The
distinguishability of one of the photons can therefore make a
significant difference in the overall quality of the state.

So far we have assumed that we know the exact behaviour of the
hidden degrees of freedom for our state.  Let us now instead
assume the experimental situation in which we can do
polarization measurements but do not know about the hidden
degrees of freedom making one of the three photons different
from the other two.  We will use a detection apparatus like the one in figure \ref{apparatus}.  The
quarter and half waveplates are set to the angles listed in the first two columns of table
\ref{waveplatesettings}.  This results in a number of detections
for each measurement outcome as listed in the last four columns.  The numbers
were generated via Monte Carlo simulation of Poisson-distributed
data arising from the density matrix in equation
\ref{theorydm}.  It was assumed that on average 10,000 three-photon states
were measured for each waveplate setting.

The set of measurement operators in
table \ref{waveplatesettings} is overcomplete, as can be verified by
explicit calculations of the
dimension of the vector space they span.  The $48$ projectors span a
space of $20$ linearly independent dimensions.  As predicted by
equation \ref{qubitdimensions}, this is the maximum number of
independent measurable operators when polarization is the only visible
degree of freedom. 
\begin{table}
\begin{tabular}{cccccc}
\hline
QWP & HWP & $3/2$ & $1/2$ & $-1/2$ & $-3/2$\\
\hline \hline
$0^\circ$ & $0^\circ$ & 3645 & 1459 & 1385 & 3586 \\
$15^\circ$ & $0^\circ$ & 2201 & 3953 & 1006 & 2703  \\
$30^\circ$ & $0^\circ$ & 275 & 7699 & 160 & 1932 \\
$45^\circ$ & $0^\circ$  & 905 & 5260 & 2904 & 904 \\
$0^\circ$ & $12.25^\circ$ & 2078 & 2042 & 3834 & 1975 \\
$15^\circ$ & $12.25^\circ$ & 2759 & 2388 & 2185 & 2673 \\
$30^\circ$ & $12.25^\circ$  & 2105 & 2693 & 4174 & 1108 \\
$45^\circ$ & $12.25^\circ$  & 420  & 6700 & 1459 & 1272 \\
$0^\circ$ & $22.5^\circ$  & 910  & 2741 & 5163 & 888 \\
$15^\circ$ & $22.5^\circ$  & 892 & 4226 & 3021 & 1899 \\
$30^\circ$ & $22.5^\circ$  & 1337 & 3838 & 3207 & 1550 \\
$45^\circ$ & $22.5^\circ$  & 1914 & 2043 & 6069 & 0 \\
\hline\hline
\end{tabular}
\caption{Simulated results of measurement of the state $\ket{30:03}
  =\frac{1}{\sqrt{2}}\left(\ket{3_H,0_V}+\ket{0_H,3_V}\right)$
  when the spatio-temporal wavepacket of one of the photons has
  a $50\%$ overlap with that of the other two.  The
  first two columns give angles for the waveplates, and the last four
  give the number of counts observed for each of the four outcomes of
  a number-resolving measurement.  These outcomes are labeled by the
  value of $m$ with the understanding that they include contributions
  from all spaces with $j\geq m$.  For example the $m=1/2$ column
  corresponds to the measurement operator $\ket{HHV}\bra{HHV}+\ket{HVH}\bra{HVH}+\ket{VHH}\bra{VHH}$.} 
\label{waveplatesettings}
\end{table}

These twenty parameters can be arranged to form an
accessible density matrix in the form of equation \ref{formofrhoacc}, with the 20 elements broken into a
16-element symmetric $j=3/2$ subspace and the remaining four elements
representing an average over the two $j=1/2$ subspaces.

Once this form is assumed for the accessible density matrix, the data
can be fit to it using maximum-likelihood fitting\cite{Kosut2004}.  To perform the fit we use the free
convex optimization package SeDumi\cite{Sturm1999} for Matlab.  In order to measure the likelihood that a
given density matrix gave rise to the dataset we calculate the
log-likelihood\cite{Kosut2004}.  The density matrix that
maximizes this function given the outcomes listed in table
\ref{waveplatesettings} is given below.    
\begin{widetext}
\begin{equation}
\rho_{\text{acc}}=\left[ 
\begin{array}{cccccccc}
{\scriptstyle 0.3626} & {\scriptstyle 0.0057+0.0033i} &
{\scriptstyle 0.0001-0.0003i} & {\scriptstyle 0.3597+0.0010i} & & & & \\
{\scriptstyle 0.0057-0.0033i} & {\scriptstyle 0.0036} &
{\scriptstyle -0.0006-0.0028i} & { \scriptstyle 0.0023-0.0040i} & &
& & \\
{\scriptstyle 0.0001+0.0003i} & {\scriptstyle -0.0006+0.0028i} &
{\scriptstyle 0.0023} & {\scriptstyle 0.0013-0.0023i} & & & & \\
{\scriptstyle 0.3597-0.0010i} & {\scriptstyle 0.0023+0.0040i} &
{\scriptstyle 0.0013+0.0023i} & {\scriptstyle 0.3601} & & & & \\
& & & & {\scriptstyle 0.0686} & {\scriptstyle -0.0322-0.0597i} & & \\
& & & & {\scriptstyle -0.0322+0.0597i} & {\scriptstyle 0.0670} & & \\
& & & & & & {\scriptstyle 0.0686} & {\scriptstyle -0.0322-0.0597i} \\
& & & & & & {\scriptstyle -0.0322+0.0597i} & {\scriptstyle 0.0670} 
\end{array}
\right]
\end{equation}
\end{widetext}
This can be seen to be very close the the density matrix in
equation \ref{theorydm}, with the difference accounted for by
the statistical noise in the measurements.  

The \emph{measured} non-zero population in the non-symmetric subspace indicates
the presence of hidden distinguishing information.  The
detection of this population would allow an experimentalist to infer the
presence of a hidden degree of freedom (in this case the
time-frequency degree of freedom) distinguishing one of the
photons from the other two.  It will also be noted that to the
extent that the photons are indistinguishable, they are indeed
in the desired state.  In other words, the errors have arisen
solely from the distinguishing information, and not, say, from
unknown polarization rotations.  This is all valuable
information useful in diagnosing problems with the experiment.

It should be emphasized that there is nothing special about the
particular waveplate settings used in this example.  The
important thing is that the resulting measurement operators fully span the
space of accessible density matrix elements.  If this is the case then
the maximum-likelihood problem is well-defined and guaranteed to
converge to the unique solution\cite{Kosut2004}.    

\section{Using the accessible density matrix to infer fundamental distinguishability}

One of the main reasons for characterizing an experimentally
generated polarization state is to substantiate claims that a
particular quantum state of light has been achieved.  For nearly
all quantum protocols only the $j=N/2$ symmetric states will be useful
since all the other states involve unwanted correlations with the
hidden degrees of freedom which, by definition, cannot be
manipulated.  Usually one makes the claim that all the photons in
the state are `indistinguishable' in the hidden degrees of freedom,
meaning that they all occupy the same hidden state.  Our technique
provides the first general method for verifying this claim.  

If,
when the accessible density matrix is measured, all the population is
found to be in the symmetric space then it must be true that the
hidden degrees of freedom are also in symmetric states.  If this were
not true then the overall state could not have the requisite bosonic
symmetry under permutation.  

If, in addition, the purity of the visible state is unity then the hidden
degrees of freedom are unentangled with the visible degrees of
freedom.  This means that all
measurements on the visible degrees of freedom
will be consistent with the photons all being in the same
single-particle hidden state.  This being the case, it makes sense to call
the photons `indistinguishable' in the conventional sense of the word.

This definition of indistinguishability is entirely consistent with
the one proposed by Liu et al.\cite{Liu2006_1,Liu2006_2} but is more
flexible because it is expressed in the density matrix formalism which
transforms in a predictable way under various operations one might
wish to perform on the state.  

\section{Conclusions}

We have outlined a procedure for measuring the state of a system
of particles spread over several experimental modes which may be
entangled with hidden degrees of freedom.  Our technique should be used to
justify claims of production of `indistinguishable' photons. It
is the most complete description of the state
possible when some degrees of freedom are hidden, and in particular it gives
a more complete description of the state than previous characterization
techniques such as those employed in
\cite{Bogdanov2004_2}, \cite{Liu2006_1} or \cite{Liu2006_2}. In addition to being
complete, this characterization also has the advantage of
producing a density matrix that can be used in the usual way to
predict the outcome of all measurements.  We expect this
method to become the standard means of characterizing states of
a fixed number of experimentally indistinguishable photons just as quantum
state tomography\cite{James2001} has become the standard means of
characterizing distinguishable photons. Indeed since the number of
accessible measurements for experimentally indistinguishable photons
only grows polynomially with the number of photons in the state, our
technique should prove useful for much larger systems of photons
than state tomography does for distinguishable photons.  

\acknowledgments{
The authors thank S.~Bartlett, S.~D'Agostino, J.~Repka, D.~Rowe and B.~Sanders for helpful discussions.
This work was supported by the Natural Science and Engineering
Research Council of Canada, the DARPA QuIST program managed by the
US AFOSR (F49620-01-1-0468), Photonics Research Ontario, the
Canadian Institute for Photonics Innovation and the Canadian
Institute for Advanced Research. RBAA is funded by the Walter C.
Sumner Foundation.  PST recognises support from iCORE, AIF and MITACS.  MWM is supported by MEC (FIS2005-03394 and Consolider-Ingenio 2010 Project "QOIT"), AGAUR SGR2005-00189, and Marie Curie RTN "EMALI".}

\bibliography{../../paper}
\end{document}